\documentclass[
]{ceurart}

\usepackage{listings}
\usepackage{floatrow}
\newfloatcommand{capbtabbox}{table}[][\FBwidth]
\usepackage{blindtext}
\begin{document}

\copyrightyear{2024}
\copyrightclause{Copyright for this paper by its authors.
  Use permitted under Creative Commons License Attribution 4.0
  International (CC BY 4.0).}

\conference{IntRS'24: Joint Workshop on Interfaces and Human Decision Making for Recommender Systems, October 18, 2024, Bari (Italy).}

\title{Bridging the Transparency Gap: Exploring Multi-Stakeholder Preferences for Targeted Advertisement Explanations}



\author[1,2]{Dina Zilbershtein}[%
orcid=,
email=zilbershtein.dina@maastrichtuniversity.nl]
\cormark[1]
\address[1]{RTL Nederland B.V., Hilversum, The Netherlands}
\address[2]{Maastricht University, Maastricht, The Netherlands}

\author[2]{Francesco Barile}[%
orcid=,
email=f.barile@maastrichtuniversity.nl]
\cormark[1]

\author[1]{Daan Odijk}[%
orcid=,
email=Daan.Odijk@rtl.nl,
]


\author[2]{Nava Tintarev}[%
orcid=,
email=n.tintarev@maastrichtuniversity.nl]

\cortext[1]{Corresponding author.}

\begin{abstract}
Limited transparency in targeted advertising on online content delivery platforms
can breed mistrust for both viewers (of the content and ads) and advertisers. This user study (n=864) explores how explanations for targeted ads can bridge this gap, fostering transparency for two of the key stakeholders. We explore participants' preferences for explanations and allow them to tailor the content and format. Acting as viewers or advertisers, participants chose which details about viewing habits and user data to include in explanations. 
Participants expressed concerns not only about the inclusion of personal data in explanations but also about the use of it in ad placing. Surprisingly, we found no significant differences in the features selected by the two groups to be included in the explanations. Furthermore, both groups showed overall high satisfaction, while ``advertisers'' perceived the explanations as significantly \textit{more transparent} than ``viewers''. Additionally, we observed significant variations in the use of personal data and the features presented in explanations between the two phases of the experiment. This study also provided insights into participants' preferences for how explanations are presented and their assumptions regarding advertising practices and data usage. This research broadens our understanding of transparent advertising practices by highlighting the unique dynamics between viewers and advertisers on online platforms, and suggesting that viewers' priorities should be considered in the process of ad placement and creation of explanations.


\end{abstract}

\begin{keywords}
Explainable Recommender Systems \sep
Advertisement Recommendations \sep
Transparency \sep
Online Behavioural Advertising
\end{keywords}

\maketitle

\section{Introduction}
\vspace{-.3\baselineskip}

As content delivery platforms (e.g. video-on-demand and music streaming platforms) increasingly utilize targeted advertisements, their influence on user engagement and content perception is undeniable \cite{freeman2022does, olney1991consumer}. This multi-stakeholder context involves three main parties: the viewers of the content (and accompanying advertisements), the platform itself and the advertisers. In this paper, we examine the priorities of two of these stakeholders: viewers and advertisers. Since targeted advertisements also impact the advertisement reach and the success of advertising campaigns, both the viewers and advertisers can benefit from a clear understanding of the process, which can help them making informed decisions and address privacy concerns \cite{dogruel2019too}. 


Within internet advertising, explanations have been touted as crucial in enhancing transparency and addressing users' concerns about data usage \cite{eslami2019user}. 
Research indicates that providing explanations for advertisements and granting users visibility and control over the information used to target them improves user experience and enhances the perception of the platform's trustworthiness \cite{barbosa2021design, kim2019seeing}. Although many online platforms now offer explanations for the advertisements, studies suggest that a significant number of users do not actually engage with these explanations \cite{ur2012smart, lee2023and}. This apparent paradox highlights the need to understand not only what users prefer in explanations, but also why they might choose to ignore them altogether. Moreover, these findings mirror the personalization privacy paradox, where transparency concerns can lead to opting out of personalization \cite{paradox}. Factors like information overload, lack of trust in explanations, or finding them irrelevant could contribute to this phenomenon \cite{ananny2018seeing}.
 
The level of detail in an explanation also remains a point of debate. While detailed explanations may seem like the ideal solution for enhancing user transparency, overly complex explanations — filled with technical jargon or excessive information — can overwhelm users. This may lead to a phenomenon known as over-reliance, where users accept the explanation at face value without fully understanding it. Some researchers argue that comprehensive explanations are necessary for transparency \cite{andreou2018investigating}, others suggest that overly detailed explanations might overwhelm users or even backfire by revealing too much about data collection practices \cite{eslami2018communicating, habib2022identifying, eslami2019user}. Finding the right balance between providing enough information and keeping explanations concise and understandable is still a key challenge in explanation design, especially with regards to harms caused by privacy violation \cite{ham2016role,najafian2023people}. 



In this work we aim to advance \textbf{transparency in targeted advertisements} by evaluating essential factors to be integrated into explanations for the two of the main stakeholders of content delivery platforms: (i) \textit{viewers} – customers of the platform consuming the ads; and (ii) \textit{advertisers} – which produce the ads that are presented to the viewers. We seek to broaden the scope of research on transparent advertising, moving beyond its conventional focus on social media platforms, investigating the perspectives of two stakeholders regarding the content and presentation of explanatory information accompanying advertisements. Our research is guided by two research questions: 
\begin{itemize}
    \item \textbf{RQ1.} Which features do viewers and advertisers prioritize for inclusion in advertisement explanations to improve transparency?
    \item \textbf{RQ2.} Which discrepancies exist, in terms of perceived transparency and satisfaction, between the explanations provided by viewers and the explanations provided by advertisers?
\end{itemize}

\section{User Study}
\vspace{-.3\baselineskip}
To answer our research questions,  we conducted a pre-registered randomized controlled trial with four between-subject factors (288 participants for each of three sessions).\footnote{The anonymized time-stamped preregistration of our hypotheses, procedure, and statistical analysis plan, can be found at the link \url{https://osf.io/w58b2/?view_only=9d649177d822474f85924efce36b0f82}} Participants were recruited using the online participant pool Prolific.\footnote{\url{https://prolific.co}} Proficiency in English and a minimum age of 18 were prerequisites for participation. Each participant could have taken part in the study only once. \footnote{All material for analyzing our results and replicating our user study (\emph{i.e.} collected dataset, analysis scripts) are not shared for anonymization, but will be made publicly available upon acceptance.} The study was approved by the ethical committee of our institution. 

We conducted our user study using a video-on-demand (VOD) platform as an example of a content delivery platform with targeted ads. This approach was chosen to evoke familiar scenarios that users frequently encounter and help them illustrate how explanations would be used. In the context of VOD platforms, specific advertising practices are employed: ads are often brand-awareness tools, with key targeting features based on demographic information and previously consumed content \cite{zhang2023survey}.
The goal of our study was to generate explanations and determine priorities regarding user-related features to be included in such explanations, for anonymous advertisement presented to hypothetical users of the VOD platform, utilizing a predefined set of features describing them. Although it is acknowledged that in real-world scenarios the content of the advertisement may impact the perception of the explanation \cite{lee2023and}, we refrained from presenting participants with an actual advertisement to avoid introducing additional confounding factors. 
While the features defining our hypothetical users do not directly correspond with individual participant, they are based on real online platform data for a realistic representation.

We apply an adapted Find-Fix-Verify crowdworker workflow \cite{bernstein2010soylent}, inspired by methodologies previously used to generate and evaluate tailored text in other contexts (e.g., personalized emotional support \cite{smith2014development,kindness2017designing}). The creation of the explanations is performed by two sets of participants, going through a process of \textbf{generation} (Find) and \textbf{revision} (Fix) of the explanations, addressing the question \textit{``Why am I seeing this ad?''}. Such participants are instructed to act as viewer or advertisers (roles are equally distributed among the participants). It is important to note that explanations are not created for the participant involved in the study, but for the hypothetical viewer from the given scenario. Finally, a third set of participants provide an \textbf{evaluation} (Verify) of the explanations generated. The explanations are assessed by participants (acting as either viewers or advertisers) with regards to perceived transparency of explanations as well as their satisfaction levels. 
We hypothesize that viewers and advertisers prioritize different features in explanations due to their distinct perspectives and objectives regarding ad content and presentation. Viewers, as emphasized in previous research \cite{wu2023slow}, tend to be more concerned with issues of privacy, control over personal data, and understanding how their information is being used to shape the ads they see. They are likely to favor explanations that minimize the use of sensitive personal data, focusing instead on how content preferences are leveraged. On the other hand, advertisers may prioritize features that align with their marketing goals, such as demographic targeting and user engagement metrics, as these are key to optimizing the effectiveness of their ads and reaching the desired audience \cite{brajnik2010review}.

Hence, we formalize the following hypotheses related to the sets of features selected to be included in the explanations (hypotheses related to \textbf{RQ1}):\newline
\begin{itemize}
    \item \textbf{H1a:} The average number of the features, chosen by viewers will differ from the average number of features, chosen by advertisers.
    \item \textbf{H1b:} The sets of features, chosen by viewers will be different from the sets of features, chosen by advertisers.
\end{itemize}

Additionally, we propose that the differing priorities of viewers and advertisers will result in variations in their evaluations, influenced by the roles of participants generating the explanations. Based on these considerations we formalize the following hypotheses related to \textbf{RQ2}:\newline

\indent \indent \textbf{There is a difference in the evaluation of satisfaction for the explanation ...:}

\begin{itemize}
    \item \textbf{H2a:} ... generated by the viewer and evaluated by the viewer versus the explanation generated by the viewer and evaluated by the advertiser.
    \item \textbf{H2b:} ... generated by the advertiser and evaluated by the advertiser versus the explanation generated by the advertiser and evaluated by the viewer.
    \item \textbf{H2c:} ... generated by the viewer and evaluated by the advertiser versus the explanation generated by the advertiser and evaluated by the advertiser.
    \item \textbf{H2d:} ... generated by the advertiser and evaluated by the viewer, and the explanation generated by the viewer and evaluated by the viewer.
    \end{itemize}
\indent \indent \textbf{There is a difference in the evaluation of transparency for the explanation ...:}
    \begin{itemize}

    \item \textbf{H2e:} ... generated by the viewer and evaluated by the viewer, and the explanation generated by the viewer and evaluated by the advertiser.
    \item \textbf{H2f:}  ... generated by the advertiser and evaluated by the advertiser, and the explanation generated by the advertiser and evaluated by the viewer.
    \item \textbf{H2g:} ... generated by the viewer and evaluated by the advertiser, and the explanation generated by the advertiser and evaluated by the advertiser.
    \item \textbf{H2h:}  ... generated by the advertiser and evaluated by the viewer, and the explanation generated by the viewer and evaluated by the viewer.
\end{itemize}

\begin{figure}[!h] 
    \caption{Scheme representation of hypotheses related to \textbf{RQ2}. V - viewer, A - advertiser.}
    \centering
    \includegraphics[width=0.5\linewidth]{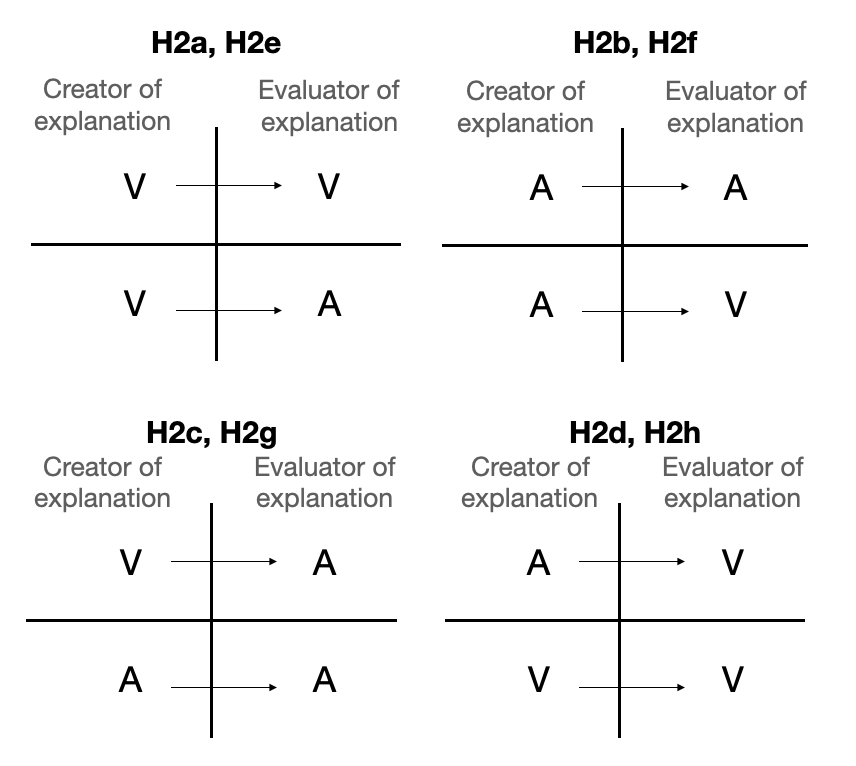}
    \label{fig:hypotheses}
\end{figure}

A visual interpretation of the hypotheses related to RQ2 is presented in Figure \ref{fig:hypotheses}.

\subsection{Procedure}

\begin{figure}[!h] 
    \caption{Scheme representation of the experimental set up. V - viewer, A - advertiser.}
    \centering
    \includegraphics[width=0.5\textwidth]{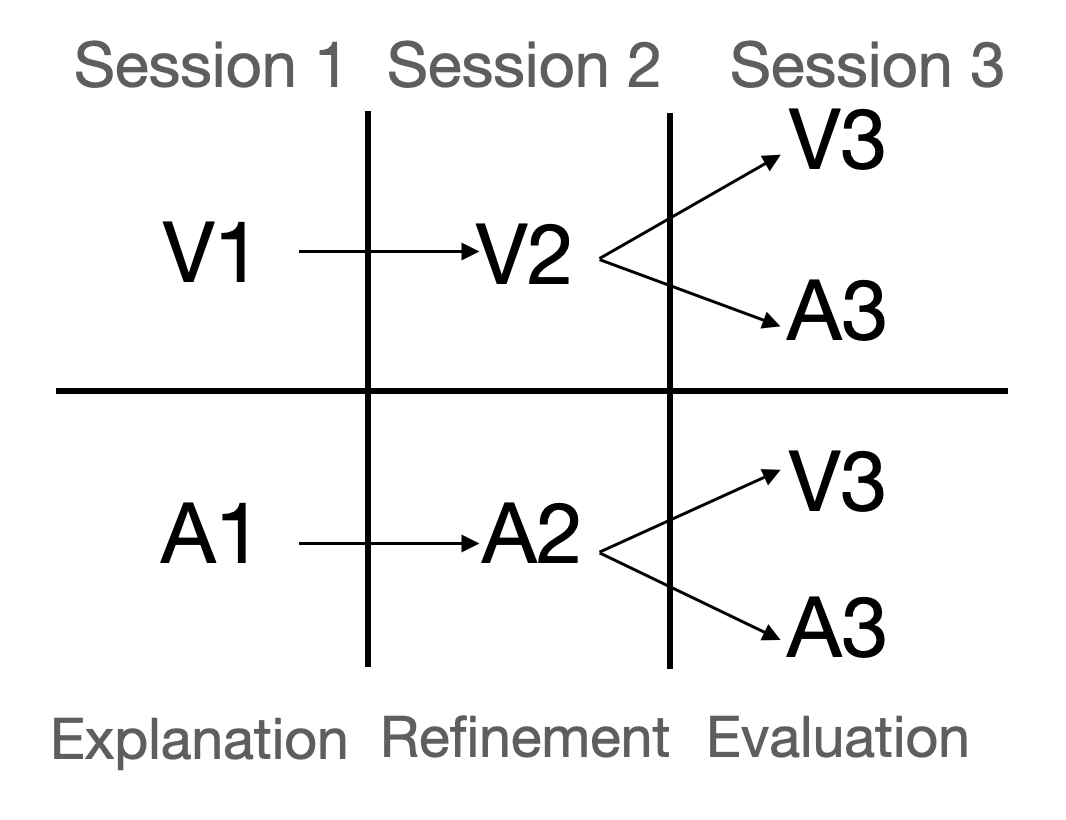}
    \label{fig:experiment}
\end{figure}

Our study consisted of three subsequent sessions, in which participants were asked to either (i) create an explanation, (ii) refine, or (iii) evaluate an explanation generated in the previous sessions. Each of the sessions consisted of a pre-survey, a task (creating/refining/evaluation of explanation) and a debrief. In all the sessions participants were assigned either the role of viewer or advertiser. Each participant engaged in only one session for a single specific scenario. A visual representation of the experimental set up is presented in Figure \ref{fig:experiment}).

\paragraph{Pre-survey.} In the first step, after obtaining informed consent, we asked participants for their gender and age group. This information was collected for statistical purposes to ensure a representative sample.

\paragraph{Role descriptions for the participants.} Participants were randomly divided into two groups: (i) \textbf{Viewers} - representing platform users who receive ads while watching content; (ii) \textbf{Advertisers} - representing advertisers, who are providing their ads to viewers through the platform. We explained both roles simply and neutrally, trying to avoid any implicit influence on their choices.
\footnote{Role descriptions can be found in the preregistration of the study.}

\paragraph{Hypothetical scenarios.} 
We defined eight scenarios, each consisting of a set of features and corresponding values, that participants could choose to incorporate into their explanations.
While real-world applications may involve a wider range of possibilities, to ensure feasibility within this research, we provided a selection of hypothetical scenarios representing diverse user groups of a VOD platform, even those relevant to smaller segments of users. \footnote{Description of the scenarios can be found in the preregistration of the study. \url{https://osf.io/w58b2/?view_only=9d649177d822474f85924efce36b0f82}}

These scenarios were derived from an analysis of data obtained from a commercial online platform, focusing on viewers subscribed to a plan inclusive of advertisements. However, it is important to note that our use of platform data analysis was strictly limited to informing feature selection. No personal user data from the platform was incorporated into the experimental scenarios. The participants were provided with a description of the scenario, as well as with the list of the features to choose from. \footnote{An example of the description, presented to the participant, can be found in the preregistration of the study.}
Variables describing the scenario were also used as features for generating explanations for advertisements. These include \textit{age group, gender, preferred content type (pref. content type), preferred genres (pref. genred), programs watched per week (programs per week), hours watched per day (hours per day), preferred days of watching (pref. type of day), preferred times of watching (pref. time of day), and preferred device (pref.device)}. This set of categorical features was used consistently across all scenarios, with varying feature values.

\paragraph{Session 1 -- Generation.} Participants were asked to generate an ad explanation based on their role and given information about a hypothetical platform user. They selected relevant features describing a user from a predefined set, which included usage patterns, consumption levels, demographics, preferred content, and device. 
Note that participants may not necessarily align with these parameters.

\paragraph{Session 2 -- Revision.} After the initial explanations were generated, a new group of participants was assigned to revise and improve them. These \textit{Revision} participants were given a role description, details about the hypothetical user receiving the advertisement, the set of features, and the previously created explanation. Using this information, their task was to refine and enhance the original explanation.

\paragraph{Session 3 -- Evaluation.} Once the explanations were refined, yet another round of participants was recruited and asked to evaluate a refined explanation in terms of perceived transparency and satisfaction, using two 7-points Likert-scale questions.


\paragraph{Debriefing.} Lastly, participants answered two open-ended questions about their preferences for advertisement explanations. First, they were asked if there was any additional valuable information they believed should be included in the explanations. They were then invited to share any further comments or feedback. 
Finally, a debriefing message was showed, with a short explanation of the objectives of the study.

\subsection{Variables}
\paragraph{Independent variables.} The combinations of creators (generation / Find) and evaluators (evaluation / Verify) of explanations was used to determine the (categorical) between-subject factors. In total, four combinations were used: (i) Viewer - Viewer; (ii) Viewer - Advertiser; (iii) Advertiser - Advertiser; and (iv) Advertiser - Viewer.

\paragraph{Dependent Variables.} \label{subsubsec:dependent_variables} For each revised explanation, we assessed two dependent variables by asking participants to rate their agreement with statements using a seven-point Likert scale ranging from ``strongly agree'' to ``strongly disagree''.
\begin{itemize}
    \item \textbf{Perception of satisfaction} (ordinal): ``The user will be satisfied with the received explanation.''
    \item \textbf{Perception of transparency} (ordinal): ``I understand why the user received an advertisement.''
\end{itemize}

\noindent \textbf{Descriptive Variables.} We also collected demographic data from participants, who had the option to opt out of providing this information.

\section{Results}
\vspace{-.3\baselineskip}

\paragraph{Participants.}

We recruited 1035 participants, with 864 of them (288 per stage: generation, revision, evaluation) passing all the attention checks. The number of required participants was determined through a power analysis on the basis of the statistical analysis plan and the number of tested hypotheses. \footnote{Analysis plan can be found in the preregistration of the study.} The average session duration was 8 minutes, exceeding the anticipated 5 minutes. To ensure fair compensation and adhere to the recommended Prolific rate of £9/hour, the remuneration for each batch was adjusted based on the observed average time spent.\footnote{\url{https://www.prolific.com/resources/how-much-should-you-pay-research-participants}.} The resulting sample exhibited a well-balanced gender distribution: 454 male, 387 female, 18 non-binary participants, 7 respondents opted not to specify their gender. Regarding age groups, the distribution was as follows: 18-24 years old (225), 25-34 years old (351), 35-44 years old (172), 45-54 years old (85), and 55 years old and above (29). Two participants preferred not to disclose their age group.

\paragraph{RQ1: differences between features selected by two groups of participants.}

\begin{figure}[!h] 
    \caption{Features selected by `Viewers' and `Advertisers' to show in the explanation for the ad.}
    \centering    
    \includegraphics[scale=0.7]{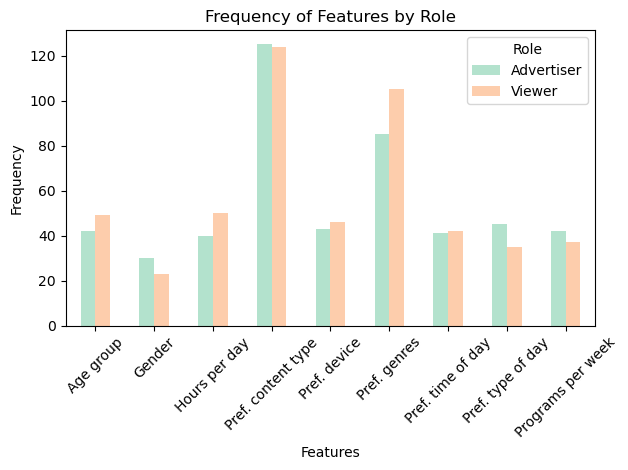}
    \label{fig:features}
\end{figure}

Our analysis found no significant difference between the count of features selected by advertiser and viewer groups (\textbf{H1a}) ($1.9$ vs. $1.8$, with standard deviations of $3.4$ and $3.5$, respectively) used in explanation (two-tail T-test, $t = -0.57$, $p = 0.56$). We also found no significant difference in the features selected by participants when comparing the \textit{Generation} and \textit{Revision} explanations (two-tail T-test, $t = 0.131$, $p = 0.896$). Additionally,  participants from both groups did not significantly differ in how often (See Fig. \ref{fig:features}) they used specific features when creating explanations (\textbf{H1b}). The analysis resulted in a Chi-Square statistic $\chi^2$ of $2.538$ ($p = 0.959$, Chi-Square Test for Homogeneity), with 8 degrees of freedom.

\begin{table}[!t]
\caption{Results of two two-way ANOVAs for the dependent variables (DVs) perception of transparency (left) and perception of satisfaction (right). Creator/Evaluator – role of the participant who created/evaluated the explanation.
}
\label{tab:anova_results}
\begin{tabular}{lll|lll}
\multicolumn{3}{l|}{Transparency}   & \multicolumn{3}{l}{  Satisfaction} \\ \hline
                  & F     & \textit{p}     &                   & F     & \textit{p}     \\ \hline
creator           & 1.915 & 0.167 & creator           & 0.302 & 0.583 \\
\textbf{evaluator} & \textbf{11.211} & \textbf{0.001} & evaluator          & 1.042         & 0.308         \\
creator:evaluator & 3.685 & 0.056 & creator:evaluator & 3.852 & 0.051
\end{tabular}%
\end{table}

\paragraph{RQ2: discrepancies in terms of perceived transparency and satisfaction between two groups of participants.}

While our analysis found no significant difference in satisfaction (\textbf{H2a-H2d}) based on who created and evaluated the explanations (viewers vs. advertisers) (see Table~\ref{tab:anova_results}), it did reveal a significant effect of the evaluator's role on the evaluation of perceived transparency (\textbf{H2e-H2h}). Advertisers perceived explanations as significantly more transparent than viewers. Their evaluations were 19\% higher (more positive) on a scale ranging from \textit{"Somewhat agree"} to \textit{"Strongly agree."}
In order to investigate the effect of the explanation creator role (advertiser vs. viewer) on perceived transparency (hypotheses \textbf{H2e \& H2f}), we conducted two separate two-tailed t-tests. The results verified hypothesis \textbf{H2e}, revealing that advertisers rated explanations created by viewers as significantly more transparent ($t = 3.547$, $df = 135.8$, $p < 0.001$). However, no significant difference was found in advertiser evaluations of transparency based on the explanation creator. Additionally, explanations generated by the participants were generally well-received by both viewers and advertisers. Around 79\% of respondents found them transparent (rated \textit{"Somewhat agree"} to \textit{"Strongly agree"}), and 58\% found them satisfactory (rated \textit{"Somewhat agree"} or higher). 




\paragraph{Qualitative analysis of the feedback from participants.} In addition to the quantitative analysis of the responses, we also examined which information participants valued in explanations. In general, they expressed that they preferred explanations that avoid detailed information about the data used for targeting. This suggests that transparency regarding targeting may raise privacy concerns for some participants, potentially outweighing the benefits of such detail: \emph{"As a user, I would have some concerns about my personal information if ad explanations had a lot of information about me"}. 
Moreover, participants mentioned that they do not want to see their demographic information in the explanations, even when they acknowledge that this data is used for targeting purposes: \emph{"It can be somewhat unsettling that the ad mentions age and gender because it can make the user feel their privacy is being violated"}. 
Respondents also acknowledged that it is hard to find a balanced approach in the creation process: \emph{"It must balance the right amount of detail with being understandable and I think it's challenging."} 

Participants generally found the presented features sufficient for explanations: \emph{"I think the most valuable information is already presented."}, \emph{"I believe that's the only information that should be used in the advertisement"}. Additionally, several participants suggested including geolocation information used on the platform.
Nevertheless, participants highly evaluated the explanations even without specific features mentioned. For example, explanations like the following were rated as high as less specific ones (see below): \emph{"You are seeing this ad, because we've tailored it to match your interests. With your penchant for action series, thrill-seeking adventures, and a sprinkle of horror, we figured you'd appreciate what we're showcasing. Given your preference for watching on your phone, we want to ensure you catch our recommendation at the ideal moments—whether it's after 6:00 PM as you wind down or before 9:00 AM to kick-start your day. Keep an eye out for our suggestion; we believe you are going to love it to keep you company during the week!"}.

A less specific example would be: \emph{"You are seeing this ad, because you are eligible in the population sample of our advertisement. Our product is directed to users with similar traits with you and similar interests."} However, too short explanations like \emph{"You are seeing this ad, because it relates to your content preferences as a user"} were evaluated low w.r.t to both transparency and satisfaction.

\subsection{Exploratory analysis.}
During the initial analysis of the participants' data, we observed variability in how they followed the instructions: the level of detail in the explanations differed, and this did not always correspond with the number of features they selected. To further explore these discrepancies, we conducted an in-depth exploratory analysis of the data, focusing on the information actually included in the text of the explanations.
\paragraph{Disparities between selected features and textual explanations.} Further analysis revealed a discrepancy between participants' stated feature selections and their actual usage in explanations in both \textit{Generate} and \textit{Revise} sessions. While instructed to incorporate specific features, participants exhibited varying levels of detail, ranging from simple feature names to in-depth descriptions. To better understand these discrepancies we examined the frequency and depth with which participants mentioned selected features and their corresponding values in their explanations. Additionally, we aimed to determine if these patterns differed between the two sessions, where participants either generated explanations from scratch or revised existing ones. 
To assess the features usage, we labeled explanations and features using a binary classification system:
\begin{itemize}
    \item \textbf{Corresponding features:} If the feature was selected and at least the name of the feature was indicated in the explanation, the label for this feature was set to 1. If the selected feature did not appear in the explanation at all, the label was 0.
    \item \textbf{Explicitly mentioned features:} If the feature was selected \emph{and the value} of this feature was mentioned in the explanation, the label for this feature was set to 1. Otherwise, it was labeled 0.
\end{itemize}

Based on these labels, we defined two scores for each explanation: (i) the \textbf{correspondence score}, which represents the ratio of corresponding features to the total number of selected features, and (ii) the \textbf{explicitness score}, which measures the proportion of explicitly mentioned features relative to the selected features. We provide examples of the labelling process and the obtained scores for three explanations in Table \ref{tab:labelling}.

\begin{table}[!t]
\caption{Examples of the labelling and related correspondence and explicitness scores.}
\label{tab:labelling}
\resizebox{\textwidth}{!}{%
\begin{tabular}{l|l|l|l|l|l|l}
\multicolumn{1}{c|}{\textbf{Explanation}}                                                                                                                                                                                                & \multicolumn{1}{c|}{\textbf{Selected features (names)}}                                                                                                   & \multicolumn{1}{c|}{\textbf{Selected features}} & \multicolumn{1}{c|}{\textbf{\begin{tabular}[c]{@{}c@{}}Corresponding\\  features\end{tabular}}} & \multicolumn{1}{c|}{\textbf{\begin{tabular}[c]{@{}c@{}}Correspondence\\ score\end{tabular}}} & \multicolumn{1}{c|}{\textbf{\begin{tabular}[c]{@{}c@{}}Explicitly \\ mentioned features\end{tabular}}} & \multicolumn{1}{c}{\textbf{\begin{tabular}[c]{@{}c@{}}Explicitness\\ score\end{tabular}}} \\ \hline
\begin{tabular}[c]{@{}l@{}}You are seeing this ad\\ because you spent\\ less than 2 hours \\ watching series per day\\ and you prefer series based\\  on the intellectual category.\end{tabular}                                         & \begin{tabular}[c]{@{}l@{}}Pref. content type,\\ Pref. genres, \\ Hours per day\end{tabular}                                                              & {[}3, 4, 6{]}                                   & {[}1, 1, 1{]}                                                                                   & 1                                                                                            & {[}1, 1, 1{]}                                                                                          & 1                                                                                         \\ \hline
\begin{tabular}[c]{@{}l@{}}You are seeing this ad, \\ because you enjoy the quality\\ of the series. Those series are\\  perfect for including them\\ in your schedule and the\\  quality is enough to \\ watch them on TV.\end{tabular} & \begin{tabular}[c]{@{}l@{}}Pref. content type,\\ Pref. genres, \\ Programs per week, \\ Hours per day, \\ Pref. type of day, \\ Pref. device\end{tabular} & {[}3, 4, 5, 6, 7, 9{]}                          & {[}1, 0, 0, 0, 0, 1{]}                                                                          & 0.33                                                                                         & {[}1, 0, 0, 0, 0, 1{]}                                                                                 & 0.33                                                                                      \\ \hline
\begin{tabular}[c]{@{}l@{}}You are seeing this ad \\ because your preferred\\ programme genres and\\ demographic makes you\\  a suitable candidate.\end{tabular}                                                                         & Age group, Pref. genres                                                                                                                                   & {[}2, 4{]}                                      & {[}1, 1{]}                                                                                      & 1                                                                                            & {[}0, 0{]}                                                                                             & 0                                                                                        
\end{tabular}%
}
\end{table}

\begin{figure}[!t] 
    \caption{Features selected/corresponding/mentioned explicitly by the participants in Session 1 (\textit{Generate}) and Session 2 (\textit{Revise}).}
    \centering    
    \includegraphics[scale=0.33]{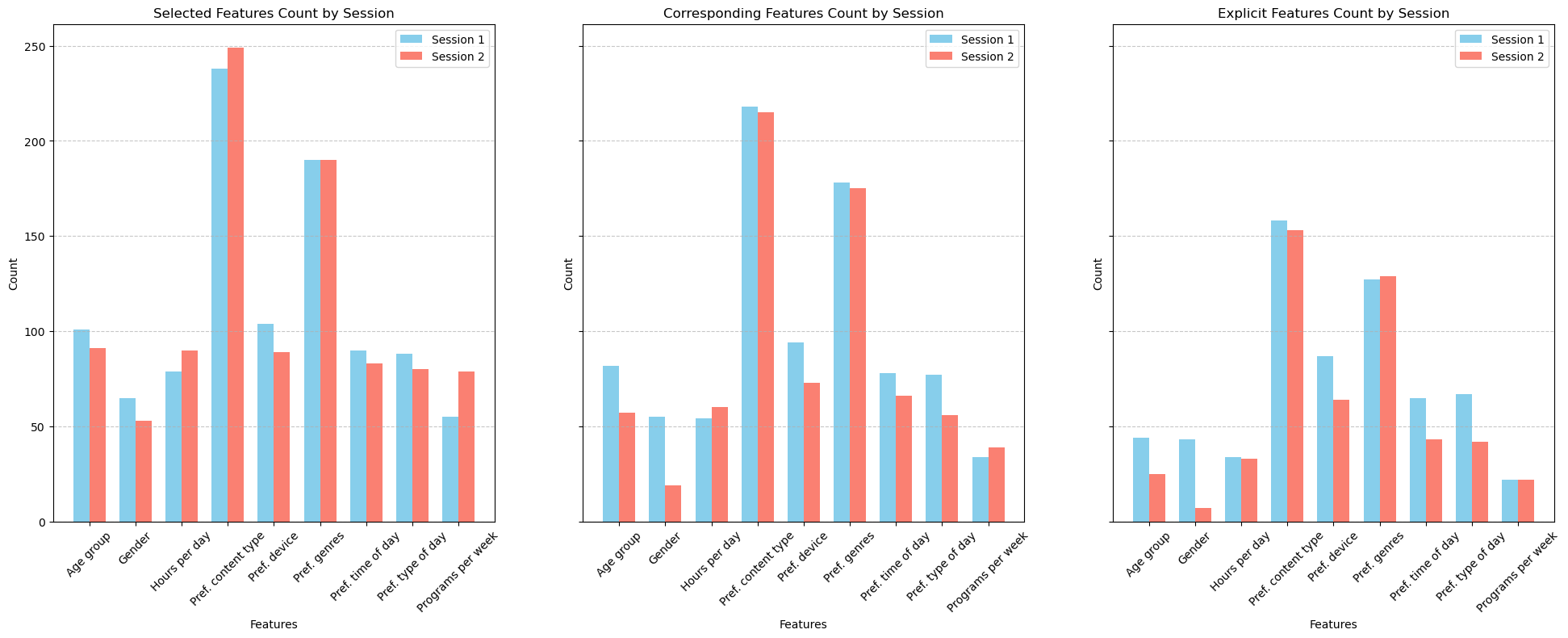}
    \label{fig:features}
\end{figure}

\begin{table}[!t]
\caption{Differences in features corresponding with the explanation between Session 1 (\textit{Generate}) and Session 2 (\textit{Revise}).}
\label{tab:feats_corr_diss}
\resizebox{\linewidth}{!}{%
\begin{tabular}{c|cccc}
\textbf{Feature}    & \textbf{Removed in S2} & \textbf{Added in S2} & \textbf{Removed in S2, \%} & \textbf{Added in S2, \%}  \\ \hline
Pref. content type        & 48                       & 45                       & 13.48                           & 12.64                                              \\
Pref. genres              & 57                       & 54                       & 16.62                           & 15.74                                          \\
Pref. device   & 56                       & 35                       & 25.45                           & 15.91                                          \\
Pref. time of day & 48                       & 36                       & 24.24                           & 18.18                                               \\
Pref. type of day & 50                       & 29                       & 27.03                           & 15.68                                                \\
Age group                 & 45                       & 20                       & 26.95                           & 11.98                                               \\
Hours per day           & 32                       & 38                       & 19.75                           & 23.46                                                \\
Programs per week       & 28                       & 33                       & 21.88                           & 25.78                                               \\
Gender              & 41                       & 5                        & 38.68                           & 4.72                                                
\end{tabular}%
}
\end{table}

We identified a discrepancy between selected features, and the features actually included in the explanations, with many features either omitted or mentioned without explicit value specification (see Fig. \ref{fig:features}). Moreover, focusing on the selected features, we can notice a variation between Session 1 (\textit{Generate}) and Session 2 (\textit{Revise}) regardless of assumed stakeholder type. While we did not observe a statistically significant difference in the overall number of features selected (corresponding or explicit), this might be explained by the fact that features were both added and removed in Session 2. For example, we can see from Table \ref{tab:feats_corr_diss} that some of the features (e.g., \textit{Gender}) were rarely mentioned (corresponded with the explanation) in Session 2, while some of them were mentioned and removed with almost the same frequency between sessions (e.g. \textit{Preferable content type}). Similar patterns were observed in the explicit mention of feature values within explanations (see Table \ref{tab:feats_expl_diss}). We also can see that features related to demographics frequently underwent removal during Session 2, with a comparatively low rate of addition. Table \ref{tab:totals_feats} demonstrates that while features such as \textit{Programs watching per week} also exhibited a relatively low rate of explicit value mentions, the discrepancy between explicit and corresponding mentions was less pronounced compared to features like \textit{Gender} and \textit{Age}. This finding aligns with our initial observations, where participants expressed in feedback their aversion to incorporating demographic information into explanations, despite acknowledging its role in targeted advertising. Nevertheless, we observed a significant difference in correspondence ($t = 3.88$, $p < 0.001$) and explicitness ($t = 3.614$, $p < 0.001$) scores between the two sessions. Both scores were higher in Session 1. Furthermore, we compared explicitness and correspondence scores and overall values between advertisers and viewers, finding no significant differences.

\begin{table}[!t]
\caption{Differences in features mentioned explicitly (with values) in the explanation between Session 1 (\textit{Generate}) and Session 2 (\textit{Revise}).}
\label{tab:feats_expl_diss}
\resizebox{\linewidth}{!}{%
\begin{tabular}{c|cccc}
\textbf{Feature}    & \textbf{Removed in S2} & \textbf{Added in S2} & \textbf{Removed in S2, \%} & \textbf{Added in S2, \%} \\ \hline
Pref. content type        & 53                       & 48                       & 17.26                           & 15.64                                                 \\
Pref. genres              & 54                       & 56                       & 18.43                           & 19.11                                                \\
Pref. device   & 55                       & 32                       & 26.7                            & 15.53                                              \\
Pref. time of day & 43                       & 21                       & 28.67                           & 14.0                                               \\
Pref type of day & 43                       & 18                       & 29.45                           & 12.33                                               \\
Hours per day           & 24                       & 23                       & 23.08                           & 22.12                                              \\
Age group                & 29                       & 10                       & 31.18                           & 10.75                                               \\
Gender              & 36                       & 0                        & 45.57                           & 0.0                                                    \\
Programs per week       & 17                       & 17                       & 23.29                           & 23.29                                               
\end{tabular}%
}
\end{table}

\begin{table}[!t]
\small
\caption{Features selected in both sessions and how often they were actually used/used with value in explanations.}
\label{tab:totals_feats}
\begin{tabular}{c|ccc}
\textbf{Feature}    & \textbf{Total Selected}  & \textbf{Corresponding Share, \%} & \textbf{Explicit Share, \%} \\ \hline
Gender              & 118                     & 62.71                        & 42.37                   \\
Age group                 & 192                     & 72.40                        & 35.94                   \\
Pref. content type        & 487                     & 88.91                        & 63.86                   \\
Pref. genres              & 380                     & 92.89                        & 67.37                   \\
Programs per week       & 134                     & 54.48                        & 32.84                   \\
Hours per day           & 169                     & 67.46                        & 39.64                   \\
Pref. type of day & 168                     & 79.17                        & 64.88                   \\
Pref. time of day & 173                     & 83.24                        & 62.43                   \\
Pref. device   & 193                     & 86.53                        & 78.24                  
\end{tabular}
\end{table}

\paragraph{Influence of the particular features on satisfaction levels.} We found a statistically significant influence of \textit{Preferable content type} and \textit{Preferable genres} on participant satisfaction with explanations ($\alpha = 0.05$). Specifically, participants expressed significantly higher satisfaction with explanations that explicitly mentioned either \textit{Preferable content type} or \textit{Preferable genres} ($t = 2.319$, $p = 0.021$). Moreover, satisfaction was further enhanced when both features were mentioned with explicit values ($t = 3.526$, $p = 0.001$). We did not observe a significant correlation between explicitness or correspondence scores and participant satisfaction levels.



    


\section{Discussion}
\vspace{-.3\baselineskip}

\paragraph{Selected features for ad explanations.} 
The quantitative hypothesis testing and the qualitative post-hoc analysis of the data offered two insights into participants' explanation preferences. First, participants generally preferred explanations that focused on information related to their content preferences. Second, explanations often did not include the actual feature values from the scenario, with an exception for the features \textit{Preferable genres} and \textit{Preferable content type}. This may be attributed to the fact that we put participants in the context of a VOD platform, main function of which is to deliver content to the users. Participants likely recognize that the platform already collects and uses this information, and they were generally comfortable with its usage. However, this practice could contrast with the real-world scenarios, in which advertiser highly prioritise demographic information in their targeting strategies \cite{boerman2017online}. 

Our exploratory analysis of the discrepancies between selected features and textual explanations further reinforced this trend. In the revision (Session 2), participants frequently omitted features or their values from explanations, even while recognizing their importance. Participants demonstrated a reluctance to include demographic information in their explanations, and during the revision phase, such information was frequently removed. 

\paragraph{Perceived transparency and satisfaction from explanations.} We did not observe a significant difference in participants' satisfaction ratings for the explanations when representing different stakeholder groups. 
However, in terms of transparency, the explanations received high ratings overall.  This suggests that explanations written in natural language may be generally well-received, regardless of whether they were created from the viewer or advertiser perspective. This finding also suggests that perceived transparency might be less sensitive than we expected to the specific features chosen or the number of features mentioned. 

\paragraph{Qualitative observations.} During the exploratory analysis, we observed a recurring theme in many explanations: participants frequently used compliments to the user in their explanations and often emphasized the user's ``uniqueness'': \textit{``you have an amazing taste'', ``you’re all about the thrill'', ``born explorer'', ``expert'', ``you are first to seek a knowledge''}. This finding also corresponds with the previous insights from participants' feedback, where they stated that \textit{``I've never felt anything good when advertisers tried to fit me in some box.''} Despite this, generated explanations exhibited biases related to the demographic information presented in the scenario. For example, explanations directed at users aged 45+ often assumed the presence of a family, while those targeting users aged 25-44 frequently implied the use of VOD platforms as a means of relaxation or stress relief after a "hard day of work".

\paragraph{Limitations of the study.}
We acknowledge that our study design, based on an online survey, has limitations. \textit{First}, to ensure experimental feasibility, we employed crowd-sourced participants to generate explanations, rather than actual advertisers. While this approach provided a scalable way to gather data, it may have introduced limitations in how well participants could authentically represent the priorities of advertisers. Despite our efforts to minimize bias in the role instructions, participants might have struggled to fully grasp the strategic concerns advertisers have, such as optimizing ad reach and targeting specific demographics. We note however that the task formulation was considered realistic and relevant by our partners at a commercial online platform (RTL Nederland B.V.).\footnote{\url{https://www.rtl.nl/}}
\textit{Second}, we did not expose respondents to the actual advertisements, which could have affected their ability to imagine the hypothetical scenario effectively. This design avoided the confounding influence of perceived relevance, but this is crucial to study in future work. While we are already working with a real platform, future studies will aim to involve real advertisers to enhance the theoretical grounding of the research and better represent the two key user groups — advertisers and viewers. By integrating more authentic stakeholder involvement, we can refine the theoretical framework and ensure that the study more accurately reflects the practical realities of both groups.
Nevertheless, this study yields valuable feedback and a comprehensive collection of human-generated explanations that illustrate participant perceptions of the advertiser's role.

\paragraph{Ethical Implications.} 
A key tension emerged from our study: user privacy concerns clashed with real-world advertising practices. Participants expressed discomfort with explanations including specific details about their habits, inclusion of demographic information, and the data collection processes behind targeted advertising. This highlights a disconnect between viewers' desire for transparency and the level of detail currently employed in ad explanations. 
While recommender systems research is exploring the trade-off between personalization benefits and privacy risks when disclosing user information \cite{najafian2023people}, the same balance needs dedicated investigation in the realm of advertisement explanations. 
This tension reflects a broader issue of user control over personal information. If certain personal information cannot be used for personalization, this likely has consequences for the relevance of advertisements. 
Our findings highlight the value of a collaborative approach to transparency. Viewers (as well as advertisers) should be involved not only in determining what information is used for advertising, but also in co-designing the explanations that inform them about these practices. By fostering such collaboration, we can bridge the gap between user expectations and advertising realities, creating transparency that respects user privacy and builds trust.

\section{Conclusion}
\vspace{-.3\baselineskip}

Explanations have been introduced in internet advertising contexts to improve transparency and minimize users' concerns about the use of their personal information \cite{eslami2019user}. However, few previous papers have investigated explanation needs in the multi-stakeholder context of a content delivery platform, a scenario that can be characterized by two of the main stakeholders, the viewers (customers of the platform) and the advertisers (who produce advertisements that are shown in the platform). As such stakeholders have different interests in the system, it is possible that they also have differing priorities regarding the information that should be included in ad explanations for content delivery platforms.

To address this potential difference, we presented a between-subject user study (N=864) investigating the preferences about the user-related features to be included in the explanations, from the point of view of both viewers and advertisers. In a crowdsourced pipeline we analyzed the explanations generated, refined, and evaluated by the two groups. The refined explanations were evaluated in terms of transparency and satisfaction. 
Surprisingly, we found no differences between advertisers and viewers in terms of the number of features they wanted to see in explanations. However, advertisers overall perceived the explanations to be more transparent. 
It was also surprising that explanations generated by ``viewers'' were well-received by ''advertisers'', and that we did not see many changes in the revision stage.
Furthermore, differently from what we could expect, the viewers found simple explanations very transparent, and they highlighted the desire to use short explanations. In contrast, they expressed privacy concerns about the use of demographic information such as age and gender in both advertisements placing and explanations for these. This contrast was further confirmed through an in-depth exploratory analysis, which showed that participants frequently omitted features they considered valuable from their own explanations, especially during the revision stage of the study. 

This work highlights the importance of considering multiple stakeholders in the design of explanations, in particular considering viewers' concerns about their personal information and explanation formulations. 
In the future, we plan to address the limitations of this study and involve real advertisers in the process of identifying the essential information to include into the explanation and how it can be adapted to the different advertising practices, involving more advanced techniques for targeting like the usage of the recommender systems.
\begin{acknowledgments}
    {This publication is part of the project ROBUST: Trustworthy AI-based Systems for Sustainable Growth with project number \\ KICH3.LTP.20.006, which is (partly) financed by the Dutch Research Council (NWO), RTL, DPG, and the Dutch Ministry of Economic Affairs and Climate Policy (EZK) under the program LTP KIC 2020-2024.  \textit{All content represents the opinion of the authors, which is not necessarily shared or endorsed by their respective employers and/or sponsors.}}
\end{acknowledgments}

\newpage
\small
\bibliography{References}

\end{document}